\documentstyle[aps,twocolumn,epsf]{revtex}
\draft
\newcommand{\myref}[1]{(\ref{#1})}
\newcommand{\figsize}{3.0 in}

\begin{document}
\title{Multifractal Scaling in the Bak-Tang-Wiesenfeld
Sandpile and Edge Events}
\author{Claudio Tebaldi$^1$, Mario De Menech$^1$, and
Attilio L. Stella$^{1,2}$}
\address{$^1$ INFM-Dipartimento di Fisica,\\
Universit\`a di Padova, I-35131 Padova, Italy\\
$^2$ The Abdus Salam ICTP, P. O. Box 563, I-34100, Trieste, Italy,\\
Sezione INFN, Universit\`a di Padova, I-35131 Padova, Italy }
\date{\today} 
\maketitle

\begin{abstract}
An analysis of moments and spectra shows that, while the
distribution of avalanche areas
obeys finite size scaling, that of toppling numbers is
universally characterized by a full, nonlinear multifractal spectrum. Rare,
large avalanches dissipating at the border influence the statistics very
sensibly. Only once they are excluded from the sample,
the conditional
toppling distribution for given area simplifies enough to show also a
well defined, multifractal scaling.
The resulting picture brings to light unsuspected, novel physics in
the model.
\bgroup
\egroup
\end{abstract}

%\vspace{1cm} 
%\narrowtext 

Finite size scaling (FSS)\cite{Fisher} is a widely adopted
framework for the description of finite, large systems near criticality.
In the last decade, after the work of Bak et al.\cite{BTW},
much attention has been devoted to a class of models in which
criticality is spontaneously generated by the dynamics itself, without the
necessity of tuning parameters. This self--organized criticality (SOC) has
been advocated as a paradigm for a wide range of phenomena, from earthquakes
to interface depinning, economics and biological evolution~\cite{SOC}. The
prototype model of SOC is the two dimensional (2D) Bak, Tang, and Wiesenfeld
sandpile (BTW)~\cite{BTW,Dhar}, which represents a system driven by a
slow external influx, dissipated at the borders through a local, nonlinear
mechanism.

In spite of its apparent simplicity and relative analytical 
tractability\cite{Dhar,Majumdar,Priezzhev-exponents,PVZ}, 
the 2D BTW resisted, so far, all
theoretical attempts to fully and exactly characterize its 
scaling\cite{Priezzhev-exponents}. 
These attempts were essentially all based on
the FSS ansatz. Numerical approaches, also based on
FSS\cite{Kadanoff}, 
led to rather scattered and sometimes contradictory numerical
results \cite{Manna,Lubeck,Paczusky}, which hardly concile with existing
theoretical conjectures. Thus, with its intriguing intractability, BTW
scaling remains a formidable challenge for nonequilibrium statistical
mechanics\cite{Dhar1} and it is very important to check if
FSS works in this context.

In this Letter we apply a new strategy of data collection and
interpretation, in order to determine to what extent the FSS ansatz can be
applied, or rather has to be modified, for a correct description of the 2D
BTW. Our results are striking and largely unexpected: while
compelling evidence is obtained that the probability distribution functions
(pdf) of some quantities obey FSS, for other magnitudes, whose
fractal dimensions can widely fluctuate within the
the nonlinear dynamics, this is definitely not
the case. Following our protocol of analysis, we demonstrate
that the well known difficulties in the description of the BTW are due to
unexpected, very nonstandard features of its dynamical behavior. In the BTW,
relations between different key quantities do not reduce to
standard power laws, as in FSS, and are substantially influenced by the
infrared cutoff given by the size of the system. The peculiar
fluctuations characterizing intermittent dissipation
at the cutoff scale, provide a dynamical mechanism
for unusual deviations from finite size, and even multifractal scaling.

We consider the BTW on a square lattice box $\Lambda $; to each site $i$ we
associate an integer height $z_{i}>0$, the number of grains. When $z_{i}$
exceeds a threshold $z_{c}=4$, site $i$ topples: $z_{i}\rightarrow z_{i}-4$,
while for the nearest neighbors $j$ of $i$, $z_{j}\rightarrow z_{j}+1$. At
the boundary less than $4$ neighbor sites are upgraded with consequent grain
dissipation. Further instabilities can be created by the first toppling. An
avalanche is the set of the $s$ topplings necessary to reach a stable system
configuration after addition of one grain ( $z_{k}\to z_{k}+1$ at some
randomly chosen $k\in \Lambda $), $a$ is the number of lattice sites
toppling at least once during the avalanche. A sequence of avalanches is
created by successive random additions. After sufficiently many grains,
thanks to dissipation at the borders, the sandpile reaches a steady state.
We analyzed  up to  $10^{8}$ avalanches in this state for
$L=128,256,512$ and 
$1024$.

The pdf for $a$ and $s$ do not reveal characteristic scales intermediate
between lattice spacing and linear pile size $L$. FSS postulates for them the
forms 
\begin{eqnarray}
P_{s}(s,L) &=&s^{-\tau _{s}}F_{s}\left( s/L^{D_s}\right)  \label{eq.1} \\
P_{a}(a,L) &=&s^{-\tau _{a}}F_{a}\left( a/L^{D_a}\right),  \nonumber
\end{eqnarray}
and usually assumes that different quantities characterizing an avalanche 
are simply
related by power laws, determined by sharply peaked conditional pdf. For
example, one expects that, given $a$, $s\sim a^{\gamma }$~\cite{Majumdar},
with $D_s=\gamma D_a$, and $D_s$ ($D_a$) representing 
the fractal dimension of $s$ ($a$).
Without sticking to the FSS form~\myref{eq.1}, scaling of 
$P_{s}$ is most generally described by the multifractal 
spectrum\cite{multifractal}: 
\begin{equation}
f\left( \alpha \right) =\frac{\log (\int_{L^{\alpha }}^{\infty }P_{s}(s,L)ds)%
}{\log (L)}.
\end{equation}
$f$ is the Legendre transform of the moment scaling function $\sigma \left(
q\right) $ defined by: 
\begin{equation}
\langle s^{q}\rangle _{L}=\int P_{s}(s,L)s^{q}ds\sim L^{\sigma \left(
q\right) },
\end{equation}
i.e. $\sigma \left( q\right) =\sup_{\alpha }[q\alpha +f(\alpha )]$.
Analogous definitions apply to the spectrum $g(\beta)$ 
($\beta =\log a/\log L$) 
and the moment exponent $\rho(q)$ of $P_{a}$. If Eqs.~\myref{eq.1}
hold, $f\left( \alpha \right) =-(\tau _{s}-1)\alpha $ for $0<\alpha <D_s$ and $f=-\infty $ for $\alpha >D_{s}$. Consistently $\sigma
\left( q\right) =D_{s}(q-\tau _{s}+1)$ for $q>\tau _{s}-1$ and $\sigma
\left( q\right) =0$ for $q<\tau _{s}-1$. Corresponding
expressions for $a$-quantities hold if FSS is valid. So, within FSS, both $f$
and $\sigma$ are piece-wise linear functions of their arguments\cite{nota}.

A reliable way to establish if Eqs.\myref{eq.1} hold, is by checking the
above linearity of $\sigma $ and $\rho$ in significant ranges of $q$.
Being extrapolated for $L\to \infty $ from finite $L$ moments, $\rho$
and $\sigma $ provide a very asymptotic characterization of the pdf's.
While for $P_{a}$ a constant gap $\Delta \rho (q)=\rho(q+1)-\rho(q)\sim
2.02\pm 0.03\sim D_a$ establishes already for $q=1$ (Table~\ref{tab:tab1}), 
for $P_{s}$, $\sigma ^{\prime }(1)\sim 2.5$, 
while $\Delta \sigma $ steadily increases
from $\Delta \sigma(1)\sim 2.70$ to $\Delta \sigma(8)\sim 2.92$. $\Delta
\sigma (1)\sim 2.7$, was also found in Ref.~\cite{Vesp}, based on large $L$
data. Thus, unlike for $P_{a}$, and in violation of FSS, for 
$P_{s}$ there is clearly no constant gap in the range $q\geq 1$.
The gap tends to rise to $\sim 3.0$ for increasing $q$. Therefore, we
should also expect $f(\alpha )>-{\infty }$ as long as $\alpha \lesssim 3.0$.

\begin{figure}[tbp]
  \centerline{
  \epsfxsize=\figsize
  \epsffile{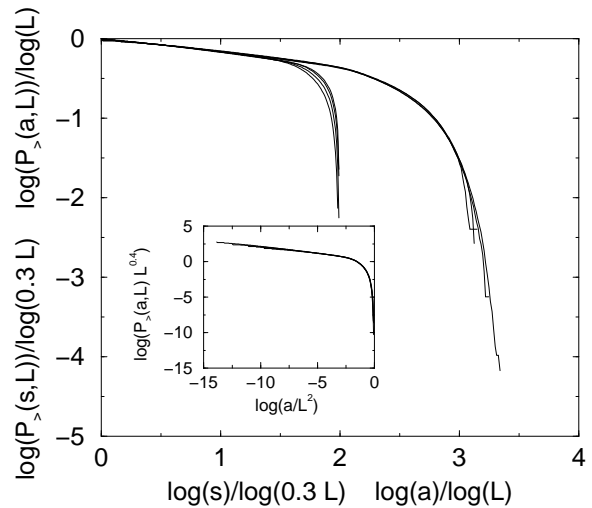}
  }
  \caption{Multifractal collapses for $g$ and $f$; a standard
  FSS collapse of $P_{>}(a,L)$ is in the inset. The subscript $>$ indicates integrated pdf.}
  \label{fig:fig1}
\end{figure}

Fig.\ref{fig:fig1}  
reports $g$ and $f$ as obtained by the data collapse technique in 
Ref.~\cite{Kadanoff}. 
The linear form $g\left( \beta \right) =-(\tau _{a}-1)\beta $ is well
verified for $\beta \lesssim 1.5$ with an estimated slope $-0.19\pm 0.01$.
In the region $\beta >1.5$, the collapse gets worse. 
One expects $D_a=2$ and, in fact, the poor $g$-collapse 
for $\beta \lesssim 2$ is
consistent with the infinite discontinuity of a FSS spectrum with
$D_a=2$: curves for various $L$ smooth out to different
degrees such discontinuity, and underestimate $g$ for 
$\beta \lesssim 2$. 
Assuming a linear $g$ in the whole domain $0\leq \beta
\leq 2$ , and $\tau _{a}=1/5$ exactly, as suggested by the estimated initial
slope, $\langle a\rangle _{L}$ 
should scale with $\rho (1)=\sup_{\beta}\left[ g(\beta )+\beta \right] 
=-2/5+2=1.6$, in nice agreement with our
determination $\rho(1)= 1.59 \pm 0.02$.
Evidence of FSS for $P_{a}$ comes also from the fact that a standard
FSS collapse ($\tau _{a}=6/5$; $D_{a}=2$) works very well 
(Fig.~\myref{fig:fig1}, inset). 
For $\alpha \lesssim 2$ the collapsed $f$ is very close to linear and
overlaps with the expected $g$ ($f\left( 2\right) \sim -0.39$). Thus, an
acceptable FSS form of $P_{s}$ should assume $\tau_{s}=\tau_{a}$, in order
to be consistent with the well collapsed, initial part of the plots. Within
such assumption, $D_s=2.5$ would be imposed by the exact result 
$\sigma (1)=2$\cite{Dhar} (we find $\sigma \left( 1\right)=1.99\pm 0.02$). 
Indeed, $\sup_{\alpha }[f(\alpha )+\alpha ]=2$ in
such case, the $\sup$ being attined at $\alpha=D_s=2.5$.
The hypothetical linear spectrum should therefore have support in 
$0\leq \alpha \leq 2.5$, and satisfy 
$f\left( 5/2\right)=-1/2$. In Ref.~\cite{demenech}, a relatively 
limited
analysis of $\sigma$ suggested $\Delta \sigma(q)\simeq 2.5$ for 
$q\geq 1$. Such constant gap would leave room to a
FSS approximation for $P_{s}$, with $\tau _{s}\simeq 6/5$ and $D_{s}\simeq
2.5$, and a linear $\sigma$ deviating from the measured one possibly
only at very low $q$'s.
According to Eq.~\myref{eq.1}, 
for such effective $P_s$ one 
would
also have $D_{s}(2-\tau _{s})\simeq 2=\sigma(1)$. However, this does not
agree with the plots which show that $f>-\infty $ at least up to $\alpha
\sim 3.0$. Indeed, curves for various $L$ collapse rather well for $\alpha
\lesssim 3.0$, and clearly suggest a support of $f$ asymptotically bounded
by $\alpha \sim 3.0$, consistent with the trend of $\Delta \sigma $ 
(Table~\ref{tab:tab1}). This bound follows also from the leftward
trend of the curves for increasing $L$
in the region $\alpha \gtrsim 3$, where collapse gets worse. An 
even approximate 
$\tau_{s}$ exponent, so extensively discussed in the last
decade, can not be simultaneously consistent with the initial slope $\sim
-1/5$ of $f$, $\Delta\sigma(\infty)\sim 3$ and $\sigma(1)=2$. 
Full consistency, in particular with $\sigma (1)=2$, can be recovered by
assuming that $f$ is indeed linear, with the same slope $-1/5$ as 
$g$, up to $\alpha =2.5$ ($f(2.5)=-1/2$), but has a nonlinear continuous
drop in the range $2.5<\alpha <3.0$. The slight underestimation by the plot
for $\alpha \sim 2.5$ ( $f(5/2)\sim -0.57$) should again be imputed
to roundoff and slower $L$-convergence in correspondence to the major
bending of $f$ \cite{Ali}.

\begin{figure}[tbp]
  \centerline{
  \epsfxsize=\figsize
  \epsffile{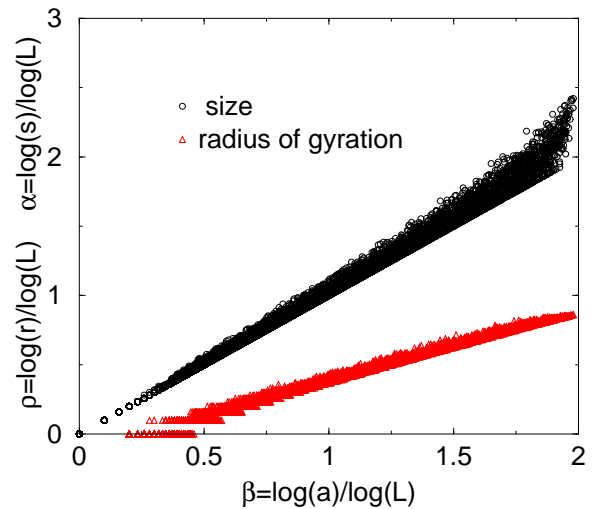}
  }
  \caption{$(\beta,\alpha)$ points for avalanches with $L=1024$ (upper);
  similar plot for $(\beta,\log r/\log L)$ (lower). The slope in the
  latter case is  $D_a^{-1}=0.50 \pm 0.01$.}
  \label{fig:fig2}
\end{figure}

The striking difference described above between $P_{s}$ and $P_{a}$ suggests
that the conditional pdf, $C$, such that 
\begin{equation}
P_{s}(s,L)=\int da\;C(s|a,L)P_{a}(a,L),
\end{equation}
should have unusual structure. Within FSS one would assume $C\sim
\delta (s-a^{\gamma })$ for $a<L^{2}$ and $L\to \infty $. Such $C$
would lead to $\gamma =(\tau _{s}-1)/(\tau _{a}-1)$. Thus, $\tau_{s}=\tau
_{a}$ would imply $\gamma=1$, while $D_{s}=2.5$ would give $\gamma
=D_{s}/D_{a}=1.25$, already contradicting FSS. In fact $C$ is a complex, broad
pdf. Its properties elucidate and confirm our conclusions for $%
P_{a} $ and $P_{s}$. Fig.~\ref{fig:fig2} reports values of 
$\alpha =\log s/\log L$ vs. $\beta =\log a/\log L$ for $L=1024$ 
avalanches. Ratios $\gamma =\alpha /\beta 
$ range between $1$ ( $s\ge a$) and $\sim 1.25$, and their spread is not
modified appreciably by sampling data for progressively larger $L$'s. For $%
\beta \lesssim 2$ also some $\alpha /\beta >1.25$ are found, which are
too rare to be displayed in Fig.~\ref{fig:fig2}. If a
relation $s \sim a^{\gamma }$ would hold, data should coalesce into a
straight line with slope $\gamma $. To the contary, points are quite spread
and form an open angle, rather than a narrow strip, as is instead the case,
e.g., for the radius of gyration, $r$, of the surface covered by the avalanche
(Fig.~\ref{fig:fig2}).

\begin{figure}[tbp]
  \centerline{
  \epsfxsize=\figsize
  \epsffile{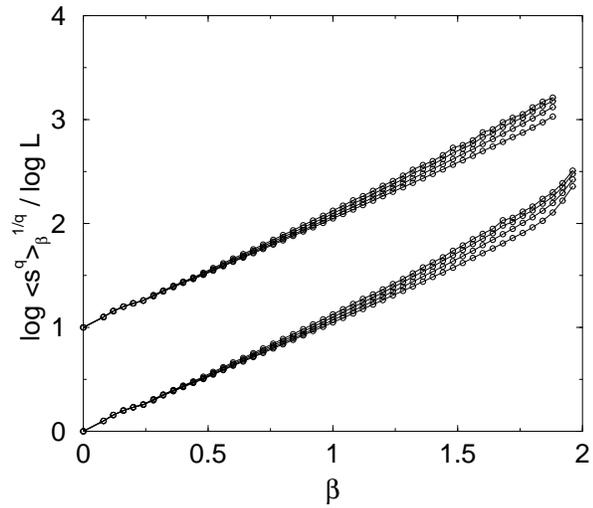}
  }
  \caption{The lower set of curves refers to all avalanches,
  while the upper one (shifted by $1$ along y axes) pertains to the
  nondissipative ones. $q=1,4,10,16$ moving upwards.}
  \label{fig:fig3}
\end{figure}

$C$-moments can be assumed to scale as 
$\langle s^{q}\rangle _{a}=
\int ds\;C(s|a,L)s^{q}\sim a^{\kappa (\beta,q)}=L^{\beta \kappa (\beta,q)}$.
One would hope
exponents like $\kappa$ to be well defined, i.e. independent of $\beta$,
for $L\to \infty$. Furthermore, with a FSS $C\propto \delta (s-a^{\gamma })$,
one should find $\beta \kappa(\beta,q)/q=\beta \gamma$ independent of
$q$. 
%This has been implicitly assumed in previous work on 
%$C$~\cite{Vesp,Christensen}. 
In Fig.\ref{fig:fig3} 
we plot $\log(\langle s^{q}\rangle_{a}^{1/q})/\log L$ 
versus $\beta $ for various $q$ and $L=1024$. The
curves, which correspond to $\beta \kappa(\beta,q)/q$ asymptotically, do not
overlap. Moreover, for each $q$ the plots have pronounced curvature,
especially for $\beta \gtrsim 1.7$. This curvature does not decrease
appreciably, or increase signalling crossovers, if one considers
progressively larger $L$'s. 
Thus, in spite of the sensible spread
of the various curves in 
Fig.\ref{fig:fig3}
(which also persists for increasing $L$), 
the complex scaling of $C$-moments can not even be classified as
multifractal. Indeed, in that case one should still require $\beta \kappa/q$
to be linear in $\beta$\cite{multifractal}. On the other hand,
the $\beta$-nonlinearity described above is essential in view
of our conclusions on $g$. Since $\langle s^{q}\rangle_{L}=\int
da\;P_{a}\left( a,L\right) \langle s^{q}\rangle _{a}$, we must have $\sigma
(q)=\sup_{\beta }[g(\beta )+\beta \kappa (q,\beta )]$. With $%
g=-\beta /5$, as argued above, the $\sup$ for $q=1$ should fall at $\beta
=2$ (Fig.~\ref{fig:fig3}), and be equal to $-2/5+2\kappa(1,2)$. Thus, one
must find $\kappa(1,2)=1.2$. If $\kappa (1,\beta )$ would remain constant
with its initial value $1.04\pm 0.02$\cite{Christensen} in the whole $\beta 
$-range, we could certainly not find $\sigma (1)=2$. Our determinations
consistently extrapolate to $\kappa(1,2)\sim 1.20$ (Fig.~\ref{fig:fig3}).

\begin{figure}[tbp]
  \centerline{
  \epsfxsize=\figsize
  \epsffile{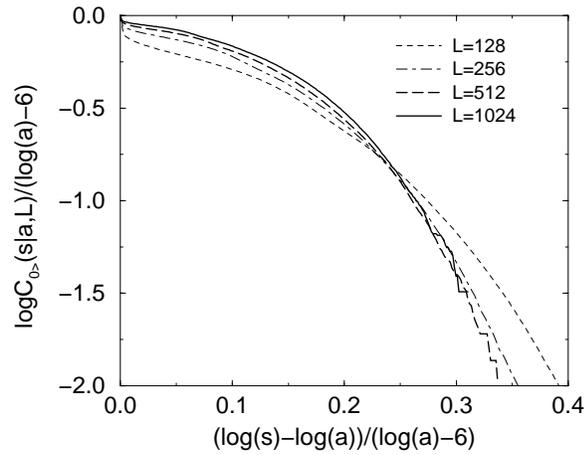}
  }
  \caption{Empirical $h(\gamma)$ at $%
  \beta =1.6$ for different L.}
  \label{fig:fig4}
\end{figure}

The $\beta $-dependence of $\kappa $ is crucial for the consistency of the
overall scaling properties and is determined by large dissipating avalanches.
These are intermittent, edge events, which guarantee grain
conservation at stationarity\cite{demenech}.
Remarkably, if only nondissipative avalanches are sampled, one obtains a
different conditional pdf, $C_{0}$, whose $\kappa_{0}$ is now
independent of $\beta$ (Fig.\ref{fig:fig3}). 
Thus, for $C_{0}$ it makes sense to discuss
a multifractal spectrum
\begin{equation}
\beta h(\gamma )=\frac{\log (\int_{L^{\beta \gamma }}^{\infty
}ds\;C_{0}(s|a,L))}{\log (L)}.
\end{equation}
As a function of $\gamma $, $h$ measures the density of points ($\alpha
=\beta \gamma ,\beta $) along a narrow vertical strip at fixed $\beta$ in
the plot correspoding to that in Fig.\ref{fig:fig3}. 
Fig.\ref{fig:fig4}
reports a data collapse of $h$-curves at $\beta =1.6$\cite{nota1}. Even if the
collapse suffers of poor $L$-asymptoticity, all curves cross nicely
at a point corresponding to $\gamma \sim 1.25$, where their trend for
increasing $L$ inverts itself.
Collapses at different $\beta$'s give very similar results. Thus, 
$\gamma \sim 1.25$ qualifies as a maximum exponent for nondissipating, bulk
avalanches. An independent, accurate determination based on rank ordering
statistics gave $\gamma=1.28\pm 0.03$.

$\gamma=5/4$ was conjectured as a unique exponent for all avalanches in 
Ref.\cite{Priezzhev-exponents}. Different values\cite{Christensen}, 
sometimes larger than $1.25$\cite{Vesp},
were conjectured in the recent literature on the basis of FSS and numerical
results. Figs. \ref{fig:fig3} and \ref{fig:fig4} show that the determination
of $\gamma$ is not even a well defined task, and can be discussed, within
a multifractal framework, only once edge avalanches
are eliminated. Indeed, for nondissipative bulk avalanches, $\gamma$
has a $\beta$-independent, broad, nonlinear spectrum, $h$, with $1$ as most
probable, and $\sim 1.25$ as extreme, most rare values.

The above analysis identifies the inadequacies of the scaling theory in 
Ref.\cite{Priezzhev-exponents}. 
Crucial to that approach was the assumption of a
relation, $n\propto a^{(2\gamma -2)/2}=a^{1/4}$, satisfied by $n$, the
maximum number of topplings in an avalanche (realized at the injection
point), and $a$, seen as the area of the first in a succession of waves \cite
{Priezzhev-exponents}. We found that also the $n$-pdf at given $a$ is broad,
and $1/4$ is only the approximate maximum value of the exponent realized by
bulk sandpile dynamics.

The constant gap $\Delta\sigma\simeq 2.5$ at high $q$ postulated in Ref.~\cite{demenech}
would have been compatible with a narrow, or even point-like spectrum for
dissipating avalanches ($\alpha \simeq 2.5$).
To the contrary, it turns
out that also dissipating avalanches are multifractal ($2.0\lesssim \alpha
\lesssim 3.0$) consistent with our more correct estimate of $%
\Delta \sigma $ at high $q$. Moreover, a frequency $\sim L^{-\zeta}$ with 
$\zeta=\frac{1}{2}$ is precisely associated to dissipating avalanches 
with $\alpha \geq 2.50$.

In summary, we showed how the long standing puzzle of 2D BTW scaling finds
a solution in a strong violation of FSS by both $P_{s}$ and $C$. While 
$P_{a}$ obeys FSS, with $D_{a}=2$ and $\tau _{a}=6/5$ as most plausible
exponents, nonlinear multifractal spectra are needed to characterize $P_{s}$
and $C_{0}$. Our results throw light on the intriguing difficulty of this
model. In spite of the several exactly known steady state properties, the
belief that the 2D BTW should be ``easily'' solvable appears unjustified.
The unusual scaling pattern discovered here, which is not found in more
simple systems, like directed sandpiles~\cite{Dhar1}, constitutes genuine
novel physics and enhances the
paradigmatic role of the BTW for statistical mechanics out of equilibrium.

The most striking feature of this pattern is the bending of the curves in
Fig.\ref{fig:fig3}, 
determined by the effect of intermittent, edge avalanches, and
essential in order to fulfill the exact constraint of local, Laplacian
conservation at stationarity ($\sigma(1)=2$). In fact the very
presence of edge avalanches allows the existence of a peculiar, bulk 
multifractal scaling of $C_{0}$.
The first moment of the $s$-pdf of nondissipative avalanches
scales $\sim L^{1.8}$, rather than
$\sim L^{2}$, as conservation imposes in the case of $P_s$. The role 
played here
by intermittent edge avalanches has striking analogies with the intermittent
phenomena in fully developed turbulence, where they cause deviation from
pure Kolmogorov scaling\cite{Frisch}. Results of the present analysis have been used most recently to elucidate quantitative connections between BTW
scaling and the notoriously difficult problem of turbulence in 3D \cite{TPB}.

We acknowledge partial support from the European Network Contract 
No. ERBFMRXCT980183. We are grateful to D. Dhar for useful criticism.

%%%%%%%%%%%%%%%%%%%%%%%%%%%%%%%%%%%%%%%%%%%%%

\begin{table}
\begin{tabular}{|c|c|c|c|c|c|c|c|}
q & 1 & 2 & 3 & 4 & 5 & 6 & 7 \\ \hline 
$\Delta \rho $ & 2.02 & 2.02 & 2.02 & 2.03 & 2.02 & 2.02 & 2.02 \\ \hline     
$\Delta \sigma $ & 2.70 & 2.83 & 2.87 & 2.91 & 2.91 & 2.92 & 2.92 \\ 
\end{tabular}
\caption{Gaps at different $q$ for $P_a$ and $P_s$. 
The estimated uncertainty is $\pm 0.03$.} 
\label{tab:tab1}
\end{table}

%%%%%%%%%%%%%%%%%%%%%%%%%%%%%%%%%%%%%% FIGURES 

\begin{references}
\bibitem{nota}  Of course we assume $F_{s}(x)\to \textrm{const.}>0$ for $x\to 0$, as
clearly suggested by the trend of $\sigma$ ( see also Ref.\cite{demenech}).
Piecewise linear spectra can also be obtained with pdf made of linear
combinations of distributions like in Eq. (1). See Ref.\cite{Ali}.
\bibitem{nota1} To get $h$ one can also first collapse
the integrated $C_{0>}(s|a,L)$ divided by $\log(a)$ for different
$L$, and then extrapolate the resulting curves for $a\to \infty$. The
collapses at fixed $a$ work remarkably well for all $a$.
\end{references}
\end{document}